\documentclass[]{epl2}

\title{Vibration Induced Non-adiabatic Geometric Phase and Energy Uncertainty of
Fermions in Graphene}
\shorttitle{Vibration induced geometric phase and energy uncertainty in graphene} 

\author{Shi-Jie Xiong\inst{1} \and Ye Xiong\inst{2}}
\shortauthor{Shi-Jie Xiong \etal}

\institute{
  \inst{1} National Laboratory of Solid State Microstructures and
Department of Physics, Nanjing University, Nanjing 210093, China\\
  \inst{2} College of Physical Science and Technology, Nanjing
Normal University, Nanjing 210097, China}

\abstract{ We investigate geometric phase of fermion states under
relative vibrations of two sublattices in graphene by solving
time-dependent Sch\"{o}dinger equation using Floquet scheme. In a
period of vibration the fermions acquire different geometric phases
depending on their momenta. There are two regions in the momentum
space: the adiabatic region where the geometric phase can be
approximated by the Berry phase and the chaotic region where the
geometric phase drastically fluctuates in changing parameters. The
energy of fermions due to vibrations shows spikes in the chaotic
region. The results suggest a possible dephasing mechanism which may
cause classical-like transport properties in graphene.}

\begin{document}

\maketitle




Due to advances of material science, the graphene, as a
two-dimensional (2D) system with one layer of carbon atoms, has been
fabricated recently. Graphene exhibits striking properties which
attracted much attention of both experimentalists and theorists
\cite{1,2,3,4,5,morozov,a1,a2,a3,a4,a5,a6,a7,a8}. Particularly, the
Dirac dispersion relation of electrons in graphene and the Fermi
level near the Dirac point lead to specific features different from
those in usual metals and semiconductors. There exists a universal
maximal resistivity, independent of their shapes and mobility.
Moreover, it is found that the weak localization is strongly
suppressed \cite{morozov} that could be attributed to a dephasing
effect similar to the phase uncertainty caused by random magnetic
field. There are several theoretical studies using different methods
addressing the unusual transport properties in graphene
\cite{a1,a2,a3,a4,a5,a6,a7,a8}.

The effect of the Berry phase acquired by fermions in cyclic motions
around the Dirac point has been noticed both theoretically
\cite{bbb} and experimentally \cite{3}. \revision{The non-adiabatic
effects and Kohn anomaly due to electron-phonon interaction have
also been addressed recently \cite{ka}.} In this paper we
investigate the geometric phase of fermions induced by periodic
vibrations between two sublattices. It is shown that in a period of
vibration the geometric phase acquired by fermions can be expressed
by the adiabatic theory only in the momentum region where the level
splitting of fermions in the evolution path is much larger than
$\hbar \omega$ with $\omega$ being the frequency. Outside this
region the adiabatic condition is not satisfied. We show that the
geometric phase, the fermion average energy, and the energy
uncertainty in the non-adiabatic region are much different from
those in the adiabatic region.

The fermion band in graphene can be described by a tight-binding
Hamiltonian with one $\pi$ orbital per site on a 2D honeycomb
lattice \cite{8a}:
\begin{equation}
  H= \sum_{\langle nn'\rangle }  t_{nn'} (a^\dag_n a_{n'} +a^\dag_{n'} a_n),
  \end{equation}
where $a^\dag_n$ ($a_n$) creates (annihilates) an electron at site
$n$, $\langle \ldots \rangle$ denotes the nearest-neighbor (NN)
sites, and $t_{nn'}$ is the NN hopping. The spin indices are not
explicitly included. On a honeycomb lattice there are two
sublattices, labeled as A and B. On a perfect lattice, all the NN
hopping integrals are the same, $t_{nn'}=t_0$. When there is a
displacement ${\bf d}$ between two sublattices, the lengths of three
NN bonds connected to a site of sublattice A become
$
   l^{(j)}=\sqrt{[d \sin (\theta - \alpha^{(j)})]^2+[l_0-d \cos
   (\theta
   -\alpha^{(j)})]^2}, \,\,
$ where $j=1,2,3$ labels the three NN bonds, $\theta$ is the angle
of ${\bf d}$ related to the $x$ axis, $l_0$ is the original bond
length, and $\alpha^{(j)} = 0, \frac{2\pi}{3}, -\frac{2\pi}{3}$ for
$j=1,2,3$, respectively, is the original azimuth angle of the $j$th
bond. For a small displacement, we can keep only the first-order
terms of $d$ and obtain $l^{(j)} \sim l_0-d \cos
(\theta-\alpha^{(j)})$. The corresponding hopping integrals become $
   t^{(j)}({\bf d})  =t_0 +\lambda d \cos
(\theta -\alpha^{(j)}), $ with $\lambda$ being the coefficient of
linear dependence of $t^{(j)}({\bf d})$ on $l^{(j)}$.

By using Bloch transformation for electron operators the original
Hamiltonian becomes
\begin{equation}
   H_0 = \sum_{\bf k} t_0 \left[ e^{{\rm i} k_x l_0 }+ 2
   e^{-{\rm i} k_x l_0/2} \cos \left( \frac{\sqrt{3}}{2} k_y l_0
   \right) \right] a^\dag_{\bf k} b_{\bf k} +{\rm H.c.}
   \end{equation}
where $a(b)_{\bf k} $ is annihilation operator of electron on
sublattice A(B) with Bloch wavevector ${\bf k}$. Expanding it to the
first order of ${\bf k}$ around two irreducible $K$ points $\left(
0, \mp \frac{4\pi }{3\sqrt{3}l_0} \right)$ in the Brillouin zone,
one has
\begin{equation}
 \label{ham}
   H_0= \frac{3}{2} \sum_{\bf k} \left[ t_0 l_0 k_y \hat{\tau}_z
    \otimes \hat{\sigma}_x -t_0l_0 k_x \hat{\textbf{1}}
    \otimes \hat{\sigma}_y \right],
   \end{equation}
where $\hat{\textbf{1}}$ and $\hat{\tau}_z$ are unit and Pauli
matrices acting on two valleys at two irreducible $K$ points, and
$\hat{\sigma}_{x,y}$ are Pauli matrices on two sublattices. The
interaction between electrons and lattice displacement ${\bf d}$ is
\begin{equation}
\label{h1}
 H_1 = \sum_{\langle nn'\rangle} (t_{nn'}({\bf d})-t_0) (a^\dag_n
a_{n'} +a^\dag_{n'} a_n),
\end{equation}
where $t_{nn'}({\bf d})$ is the hopping integral between sites $n$
and $n'$ under displacement ${\bf d}$.
Keeping only the terms of the first order of ${\bf d}$, the
interaction Hamiltonian becomes
\begin{equation}
   H_1 = \frac{3\lambda }{2} \sum_{\bf k} (
 d_x \hat{\textbf{1}} \otimes \hat{\sigma}_x + d_y \hat{\tau_z} \otimes \hat{\sigma}_y
 ).
   \end{equation}
It is similar to the expression previously derived based on a
valence-force-field model \cite{su}. We only consider in-plane
optical modes at the long wavelength limit, as the modes with short
wave lengths induce large momentum transfer which causes the
electron states out of the Dirac fermion region, the vertical
optical modes only quadratically couple with the fermions, and the
coupling between acoustic modes and electrons is even more weak as
the nearest-neighbor bond lengths are almost unchanged.
We denote the in-plane vibrations of the relative coordinates as $
d_x= \frac{\eta_x}{\lambda} \cos (\omega t +\alpha_x)$ and $ d_y
=\frac{\eta_y}{\lambda} \cos (\omega t + \alpha_y)$, where
$\frac{\eta_{x(y)}}{\lambda}$ and $\alpha_{x(y)}$ are amplitude and
initial phase of the vibration in the $x(y)$ direction,
respectively.

The Hamiltonian for electrons becomes time-dependent:
\begin{equation}
   H(t) = \frac{3}{2} \sum_{\bf k} \left[ (t_0 l_0 k_y \hat{\tau}_z +
   \eta_x  \cos (\omega t +\alpha_x) \hat{\textbf{1}}) \otimes \hat{\sigma}_x -(t_0 l_0 k_x \hat{\textbf{1}}
   -\eta_y \cos (\omega t+\alpha_y) \hat{\tau_z}) \otimes \hat{\sigma}_y
   \right].
   \end{equation}
For a given valley labeled by $+ $ or $-$, the Hamiltonian is:
\begin{equation}
   H_\pm (t)= \frac{3}{2} \sum_{\bf k} \left\{ t_0 l_0 (\pm k_y
   \hat{\sigma}_x -k_x \hat{\sigma}_y) + \eta_x  \cos (\omega t +\alpha_x)
   \hat{\sigma}_x \pm  \eta_y\cos(\omega t+\alpha_y) \hat{\sigma}_y  \right\}.
   \end{equation}
\begin{figure}
\onefigure{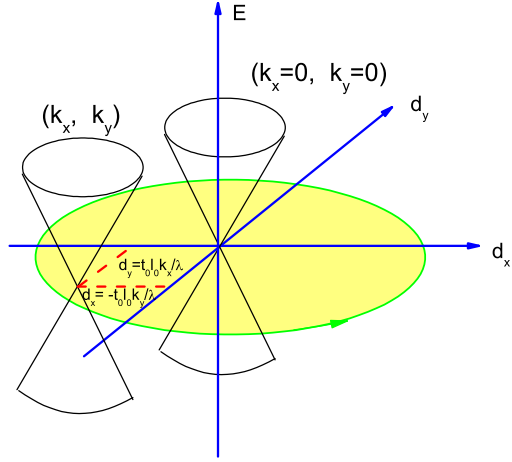}
\caption {(Color online) $E^{(+)}_{1,2}$ as
functions of ${\bf d}$ for different values of ${\bf k}$. Green
circle shows the evolution path in the vibrations. } \label{fig1}
\end{figure}
At the adiabatic limit, the instantaneous eigenenergies at time $t$
can be obtained by diagonalizing $H_\pm (t)$:
   \begin{equation}
     \label{ener}
   E^{(\pm)}_{m}(t) = (-1)^m\, \frac{ 3}{2} \left| t_0 l_0 (\pm k_y+{\rm i}k_x) +
   \eta_x \cos(\omega t+\alpha_x) \mp {\rm i} \eta_y \cos( \omega t +\alpha_y) \right|,
   \end{equation}
where $m=1$ and $m=2$ correspond to bands below and above the Dirac
point, respectively. For given $(k_x,k_y)$, these instantaneous
eigenenergies exhibit conic dependence on parameters $d_x$ and
$d_y$, as shown in Fig. 1. The diabolical point, where the poles of
two cones coincide, is determined by $(d_x=\mp
t_0k_yl_0/\lambda,\,\, d_y=\pm t_0 k_xl_0/\lambda)$, where plus and
minus signs refer to two valleys. The vibrations of $d_x$ and $d_y$
result in circular motion with an elliptic track in the $d_x-d_y$
plane. From the Berry theorem, a cyclic motion along a close track
in a 2D parameter space can cause a Berry phase of $\pm \pi$ in a
wavefunction whose instantaneous eigenenergy has a diabolical point
enclosed in this circle \cite{berry}. Since the position of the
diabolic point in the $d_x-d_y$ plane is determined by $(k_x,k_y)$,
the Berry phase acquired by the fermions with different momenta are
different: It is $\pm \pi$ for the fermions whose diabolic point
$(d_x=\mp t_0k_yl_0/\lambda,\,\, d_y=\pm t_0 k_xl_0 /\lambda)$ is
enclosed in the track, while it is zero for the states with the
diabolic point in the $d_x-d_y$ parameter space outside the track.

The above conclusion about the Berry phase is valid only in the
adiabatic condition, i.e., the energy difference between two bands
is always much larger than $\hbar \omega$ in the track. There exist
fermion states for which the adiabatic condition can not be
satisfied. In this case we have to investigate the geometric phase
from solutions of time-dependent Hamiltonian. In a periodic
time-dependent Hamiltonians the wavefunctions can be expressed in
the Floquet form \cite{floquet,fl}. This scheme has been used to
solve the states in a laser field in calculation of the harmonics
generation in graphene \cite{ak} and in nanotube \cite{hsu}. In
periodic vibrations of $d_x$ and $d_y$ the electron wavefunction can
be written as
\begin{equation}
  \psi^\pm (l,{\bf k},t) = e^{-{\rm i} \frac{\epsilon^\pm (l,{\bf k})}{\hbar}
   t} u^\pm (l,{\bf
  k}, t),
  \end{equation}
where $l$ is an index of electron states for given ${\bf k}$, and
$u^\pm(l,{\bf k}, t)$ is a periodic function of $t$ with period $T=
\frac{2\pi}{\omega}$. Performing Fourier transformation for
$u^\pm(l, {\bf k},t)$ with respect to $t$, one has
\begin{equation}
   u^\pm(l,{\bf k},t) = \sum_n \left( \begin{array}{c} A^\pm_n (l,{\bf k}) \\
   B^\pm_n (l,{\bf k}) \end{array} \right) e^{-{\rm i} n \omega t},
   \end{equation}
where $A^\pm_n$ and $B^\pm_n$ are components on two sublattices.
Substituting the Floquet wave function into the time-dependent
Schr\"{o}dinger equation $ [H_\pm (t) -{\rm i}\hbar
\partial_t ] \psi^\pm (l, {\bf k}, t) =0 $, we obtain a set of linear
homogeneous equations for the components
\begin{equation}
 \label{eq1}
  \frac{3t_0l_0}{2} (\pm k_y +{\rm i}  k_x) B^\pm_n + \frac{3}{4}
  (e^{{\rm i}\alpha_x}\eta_x \mp {\rm i} e^{{\rm i}\alpha_y}\eta_y) B^\pm_{n- 1}
  + \frac{3}{4}
  (e^{-{\rm i}\alpha_x}\eta_x \mp {\rm i} e^{-{\rm i}\alpha_y}\eta_y) B^\pm_{n+ 1}=
  (\epsilon^\pm + n\hbar \omega ) A^\pm_n,
\end{equation}
\begin{equation}
 \label{eq2}
 \frac{3 t_0l_0}{2} (\pm k_y -{\rm i}  k_x) A^\pm_n + \frac{3}{4}
  (e^{{\rm i}\alpha_x}\eta_x \pm {\rm i} e^{{\rm i}\alpha_y}\eta_y) A^\pm_{n- 1} + \frac{3}{4}
  (e^{-{\rm i}\alpha_x}\eta_x \pm {\rm i} e^{-{\rm i}\alpha_y}\eta_y) A^\pm_{n+ 1}=
  (\epsilon^\pm + n\hbar\omega ) B^\pm_n.
\end{equation}
From the requirement of nonzero solutions, for given ${\bf k}$ one
can solve the discrete quasienergies $\epsilon^{\pm}(l,{\bf k})$ and
the corresponding Floquet states $\psi^\pm(l,{\bf k},t)$. It is
noteworthy that two Floquet states whose quasienergies differ by $n
\hbar\omega$ with $n$ being an integer are physically equivalent
states \cite{fl}. So the quasienergies of physically different
Floquet states can be reduced into region
$-\frac{\hbar\omega}{2}\leq\epsilon^\pm \leq \frac{\hbar\omega}{2}$.

In a period the wavefunction $\psi^\pm (l,{\bf k},t=T)=e^{i \phi^\pm
(l,{\bf k})} \psi^\pm (l,{\bf k},t=0)$ acquires a phase
$\phi^\pm(l,{\bf k})$. It consists of two parts:
$
\phi^\pm(l,{\bf k})= \alpha^\pm(l,{\bf k})+\beta^\pm (l,{\bf k}),
$
where
$
\alpha^\pm (l, {\bf k}) = -\frac{\bar{E}^\pm (l,{\bf k}) T}{\hbar}
$
is the dynamical phase with
 $
 \bar{E}^\pm (l,{\bf k}) = \frac{1}{T} \int_{0}^{T}dt\langle\psi^\pm(l,{\bf k},t)|H_\pm
(t)|\psi^\pm(l,{\bf k},t) \rangle,
$
and
$
\beta^\pm(l,{\bf k})=\phi^\pm(l,{\bf k}) -\alpha^\pm(l,{\bf k})
$
is the geometrical phase \cite{aa}. Under the periodic vibrations
the Floquet state $\psi^\pm (l,{\bf k},t)$ is stationary one which
returns to the initial state other than phase $\phi^\pm(l,{\bf k})=
-\frac{2\pi \epsilon^\pm (l,{\bf k})}{\hbar \omega}$ after a period
of evolution. At the same time, the energy is no longer a good
quantum number and the average energy of stationary state $\psi^\pm
(l,{\bf k},t)$ can be calculated as
\begin{equation}
 \bar{E}^\pm (l,{\bf k}) = \epsilon^\pm (l,{\bf k}) + \hbar \sum_n
 n \omega \left( |A^\pm_n (l,{\bf k})|^2 +|B^\pm_n (l,{\bf
 k})|^2\right).
 \end{equation}
From this one obtains the geometric phase
\begin{equation}
 \beta^\pm (l,{\bf k}) =  2\pi \sum_n
 n  \left( |A^\pm_n (l,{\bf k})|^2 +|B^\pm_n (l,{\bf
 k})|^2\right).
 \end{equation}
The energy uncertainty caused by the vibration can be specified by
the standard variance
$$
 \Delta^\pm (l,{\bf k}) \equiv \langle E^2 \rangle - \langle E
  \rangle^2
$$
\begin{equation}
 \label{delta1}
  = \hbar^2\omega^2 \left\{ \sum_n n^2 \left( |A^\pm_n(l,
  {\bf k})|^2 +|B^\pm_n(l,
  {\bf k})|^2 \right) - \left[ \sum_n n \left( |A^\pm_n(l,
  {\bf k})|^2 +|B^\pm_n(l,
  {\bf k})|^2 \right) \right]^2 \right\}.
  \end{equation}

From the superposition principle any time-dependent state can be
expressed as a linear combination of the Floquet states. But this
linear combination is usually not a stationary state, i.e., the
state could not return to its initial one with only a phase
difference after a period. For such states the geometric phase can
not be defined. So in this paper we only consider the geometric
phase for the Floquet states. Since for given valley and given
momentum there are only two unknowns having the same phonon number
$n$ in Eqs. (\ref{eq1}) and (\ref{eq2}), the number of quasienergies
within range of $\left[ -\frac{\hbar \omega}{2}, \frac{\hbar
\omega}{2} \right] $ is 2, corresponding to the lower and upper
bands of the Dirac fermions with labels $l=1$ and $l=2$,
respectively.

\begin{figure}
\onefigure[width=6cm]{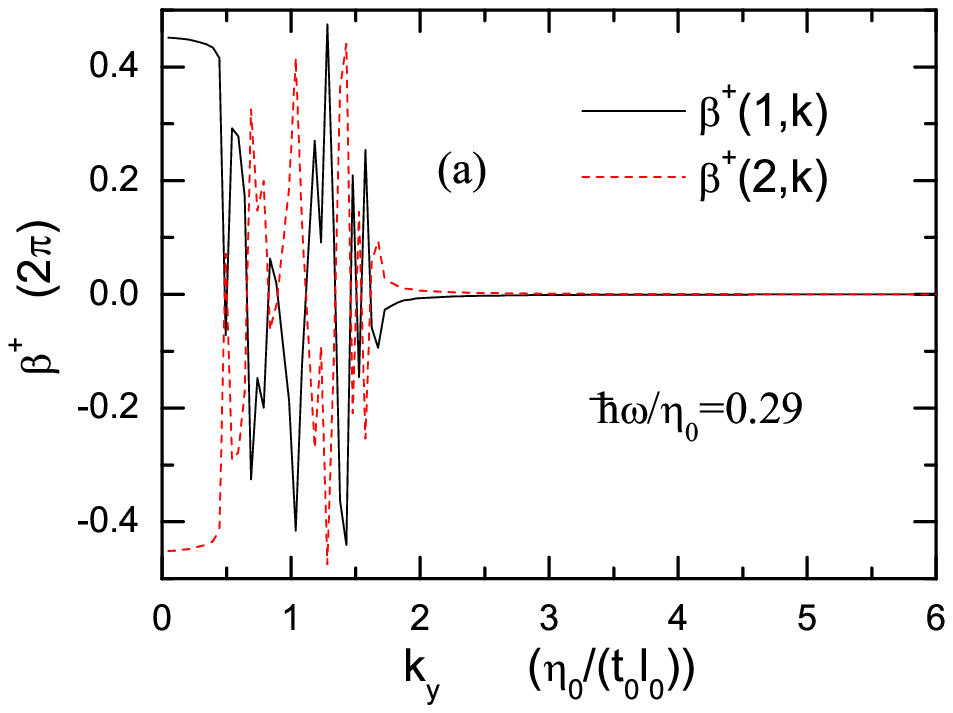}\onefigure[width=6cm]{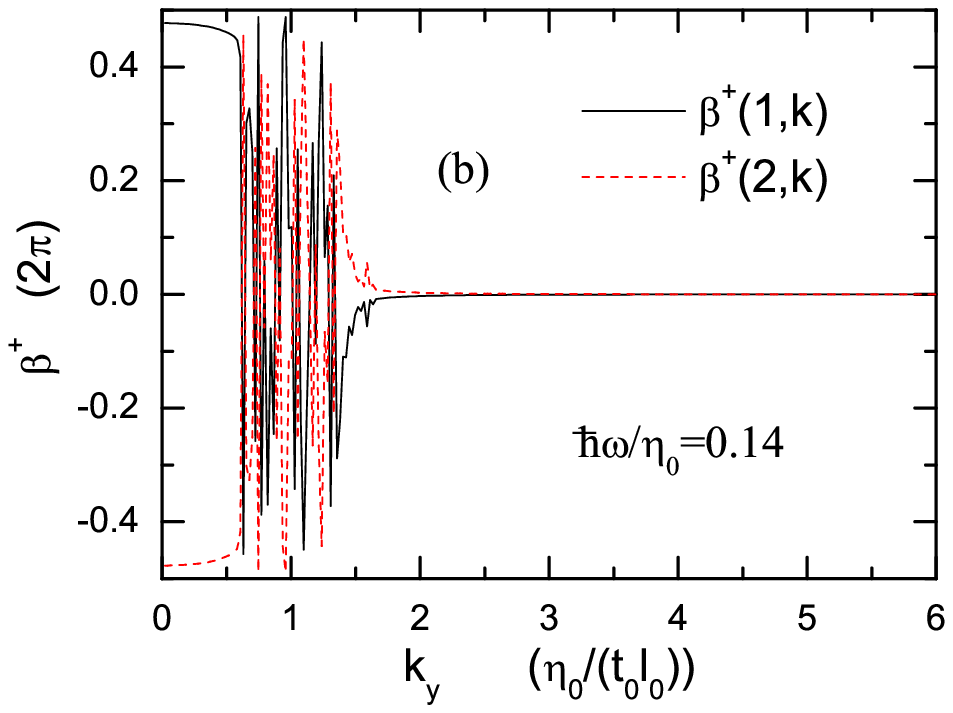}
\onefigure[width=6cm]{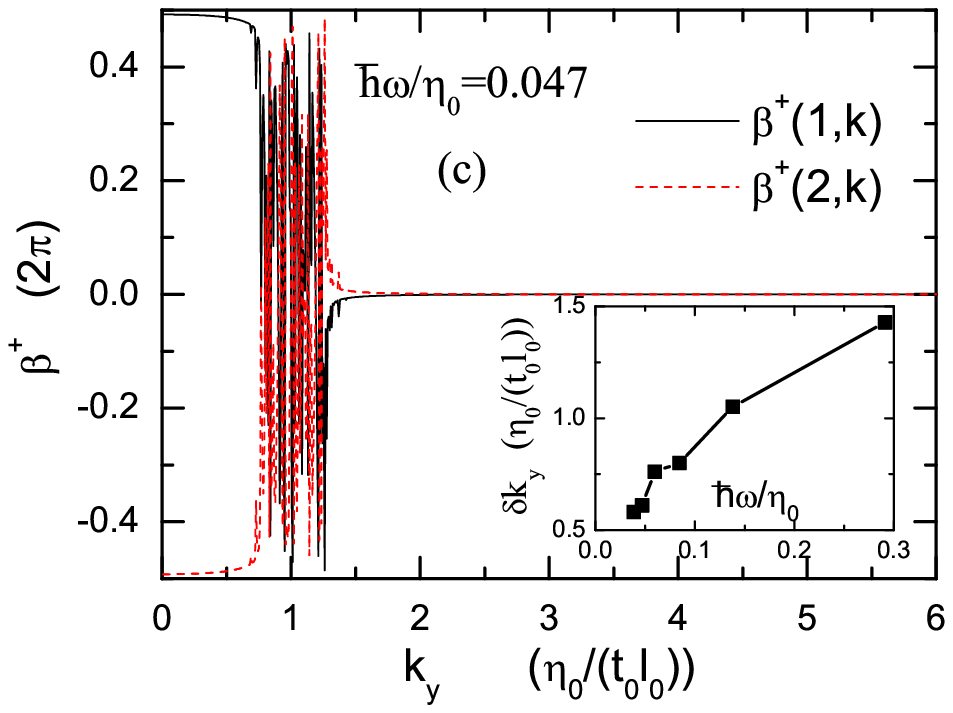} \caption{(Color online) Geometric
phase acquired in a period of vibrations by Dirac fermions in lower
and upper bands of valley ``$+$" as a function of the fermion
momentum. The parameters for vibrations are: $\eta_x=\eta_y=\eta_0$,
$\alpha_x=0$, and $\alpha_y = -\frac{\pi}{2}$. $k_x=0$ and units of
$k_y$ are set to be $\frac{\eta_0}{t_0l_0}$ so that the border
between the $\pm \pi$ Berry phase and the zero Berry phase is at
$k_y=1$. The inset of (c) shows the width of the chaotic region
$\delta k_y$ as a function of the frequency. } \label{fig2}
\end{figure}

Now we begin to investigate the properties of stationary states in
the periodic vibrations. We are interested in: (i) the geometric
phases acquired by fermion states with various momenta in a period
of vibrations; (ii) the deviation of the average energy from the
Dirac dispersion relation and the energy uncertainty caused by the
vibration. From (i) we can shed some light on the phase variation of
Dirac fermions. From (ii) we can see the essential effect of
vibrations on the basic energy spectrum.

In Fig. 2 we plot the geometric phase acquired by fermion states in
a period of vibrations versus the fermion momentum. There is a
$\frac{\pi}{2}$ phase difference between the vibrations in the $x$
and $y$ directions and their amplitudes are the same, i.e.,
$\eta_x=\eta_y=\eta_0$. So this is a circular vibration of the
relative coordinates of two sublattices as shown in Fig. 1.
According to the Berry theorem, the adiabatic Berry phase is $\pm
\pi$ if $|{\bf k}| < \frac{\eta_0}{t_0l_0}$ and is zero if $|{\bf
k}| > \frac{\eta_0}{t_0l_0}$. We note that from the calculations
beyond the adiabatic approximation the obtained geometric phase is
still roughly equal to the Berry phase: for $|{\bf k}| \ll
\frac{\eta_0}{t_0l_0}$ it is nearly $\pm \pi$ and for $|{\bf k}| \gg
\frac{\eta_0}{t_0l_0}$ it is almost zero. However, now there appears
a ``chaotic" region around the border $|{\bf k}| =
\frac{\eta_0}{t_0l_0}$ where the geometric phase randomly oscillates
in changing the momentum. This is a direct consequence of the
non-adiabaticity, as in this region the adiabatic condition, that
the energy difference between two bands is much larger than $\hbar
\omega$ in the evolution path, is not satisfied. As can be seen from
the inset of Fig. 2(c), the width of this chaotic region is
increased by increasing the frequency. The chaotic nature of the
geometric phase reflects the phase uncertainty of the single-fermion
states in this momentum region during vibrations. Such vibrations
may exist in the graphene due to thermal excitations, or due to the
zero-point fluctuations. Especially, the Berry phase effect may
produce Born-Huang vibronic centrifugal term (see below) which may
cause nonzero vibrations even at very low temperature. So such a
fermion-lattice interaction can play roles of a dephasing mechanism
for the single-fermion states. We also note that in the whole
region, even including the chaotic region, the geometric phases
acquired by fermions of the same momentum in two bands are opposite
to each other. This phase compensation effect of two bands implies
that this dephasing mechanism has no effect if the fermions in two
bands are paired. The exchange of two bands leads to the sign
reverse of the geometric phase. From the symmetry shown in Fig. 1 we
can see that an equivalent operator which can also cause the sign
reverse of the geometric phase is the reverse of the direction of
the circular vibration from clockwise to counterclockwise or vice
versa.

\begin{figure}[ht]
\onefigure[width=9cm]{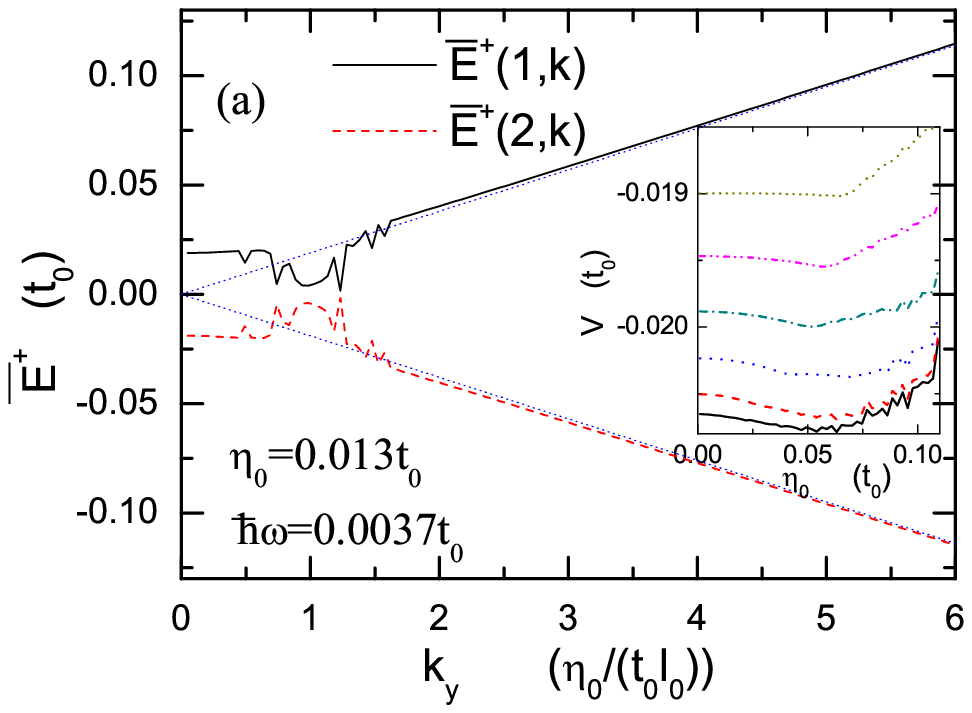}\onefigure[width=9cm]{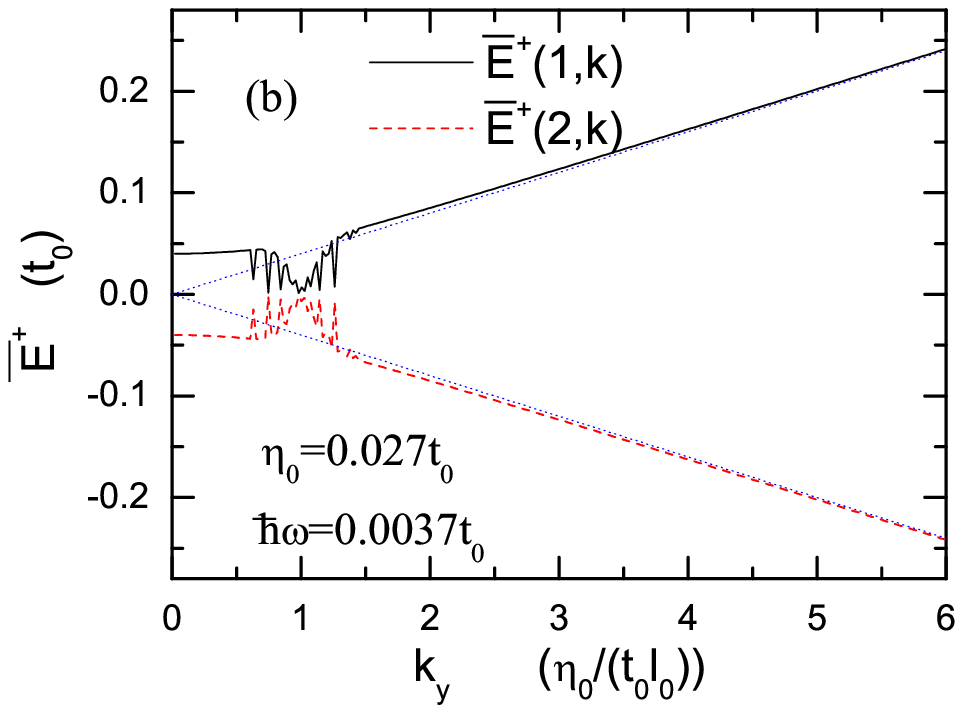}
\caption{(Color online) Average energy of fermions in lower and
upper bands of valley ``$+$" as a function of the fermion momentum.
The parameters for vibrations are $\eta_x=\eta_y=\eta_0$,
$\alpha_x=0$, and $\alpha_y = -\frac{\pi}{2}$. $k_x=0$. The blue
dotted lines represent eigenenergies without vibrations. Inset in
(a): The effective potential of vibrations as a function of rescaled
displacement $\eta_0$ for different carrier densities. For curves
from bottom to top the extra electron densities above the Dirac
point are: 0, 0.00054, 0.00108, 0.00162, 0.00216, 0.0027 per atom.
$\hbar \omega = 0.0037t_0$ and $g=0.25/t_0$.} \label{fig3}
\end{figure}

Another significant effect of vibrations is the change of fermion
energies. As the single-particle energy is not a good quantum
number, in Fig. 3 we plot the average energy of single-fermion
states as a function of the momentum. From the comparison with the
Dirac dispersion relation and with Fig. 2, the energy difference
between two bands is enlarged near the Dirac point and for $|{\bf
k}| \gg \frac{\eta_0}{t_0l_0}$, where the geometric phase is near
its adiabatic value, $\pm \pi$ or zero, while this energy difference
shrinks in the chaotic region where the adiabatic approximation can
not be used. Such opposite behaviors in these two regions imply that
the failure of the adiabatic theory in the chaotic region has much
more profound meaning than reflected from the values of geometric
phase. Physically, in the chaotic region the oscillations of
electrons between two bands are not able to follow the vibrations,
this causes the loss of distinguishability of the two bands and the
resultant states trend to take average energies in between, leading
to a smaller energy spacing. On the contrary, in a quantum
description the coupling of two states always enlarges their energy
spacing. So one may expect that the chaotic region corresponds to a
classical-like behavior and has a maximum uncertainty of
single-fermion energies.

The total average energy of fermions $V_e$ is then also changed and
becomes dependent on the vibration amplitude $\eta_0$. In the inset
of Fig. 3(a) we plot the effective potential $V(\eta_0)= \frac{1}{2}
g \eta_0^2 +V_e(\eta_0)$ where the first term is the ordinary
elastic potential. The minimum is shifted from the center
(Born-Huang vibronic centrifugal term), leading to nonzero
vibrations even at very low temperature.

We calculate the standard variance of Eq. (\ref{delta1}) and show
the results in Fig. 4. As expected, the standard variance exhibits
spikes in the chaotic region, and the energy uncertainty increases
by increasing the amplitudes of vibrations. Except for the spikes in
the chaotic region, the standard variance globally increases by
increasing the momentum. Especially, the energy uncertainty becomes
zero at the Dirac point for all the investigated vibration
amplitudes. This originates from the fact that the the energy
uncertainty is due to the oscillation of the fermion states between
two bands, but at the Dirac point the two bands coincide, leading to
a zero amplitude of the oscillation. This structure, including the
spikes and the globally increasing background with the momentum, may
catch major features of linewidth distribution obtained from the
angle-resolved photoemission spectroscopy in graphene \cite{spe}.
Spikes also appear in the average energy of electrons in quasi-1D
nanotube under periodic laser field \cite{hsu}, but in the present
case the spikes occur in 2D momentum space originating from specific
electron-phonon interaction in graphene and are closely related to
the geometric phase.

\begin{figure}[ht]
\onefigure{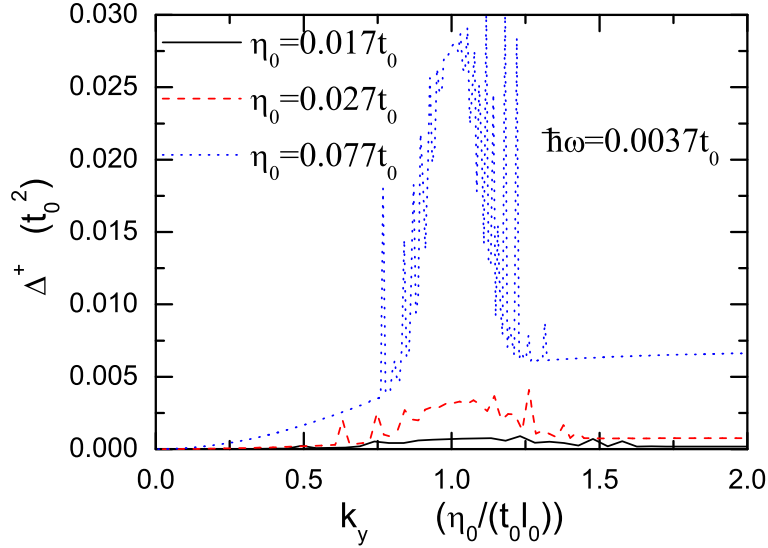} \caption{(Color online) Standard variance of
single-fermion energy of valley ``$+$" as a function of the fermion
momentum. The parameters for vibrations are $\eta_x=\eta_y=\eta_0$,
$\alpha_x=0$, and $\alpha_y = -\frac{\pi}{2}$. $k_x=0$. The standard
variance is the same for lower and upper bands. } \label{fig4}
\end{figure}

The evolution path plays a crucial role on the distribution of
adiabatic and chaotic regions in the momentum space. To show this in
Fig. 5 we plot the geometric phase as a function of $k_x$ and $k_y$
for an elliptic path. The chaotic region forms an orbicular area
with a nearly fixed width along a loop defined by $k_x
=\frac{\lambda d_y(t)}{t_0l_0}$ and $k_y =-\frac{\lambda
d_x(t)}{t_0l_0}$. The width depends on $\omega$ as shown in the
inset of Fig. 2(c). As a result, the adiabatic region with a $\pm
\pi$ Berry phase is compressed by reducing the short axis of the
ellipse, but the chaotic region with random geometric phases can
exist even for a linear vibration.

\begin{figure}[ht]
\onefigure{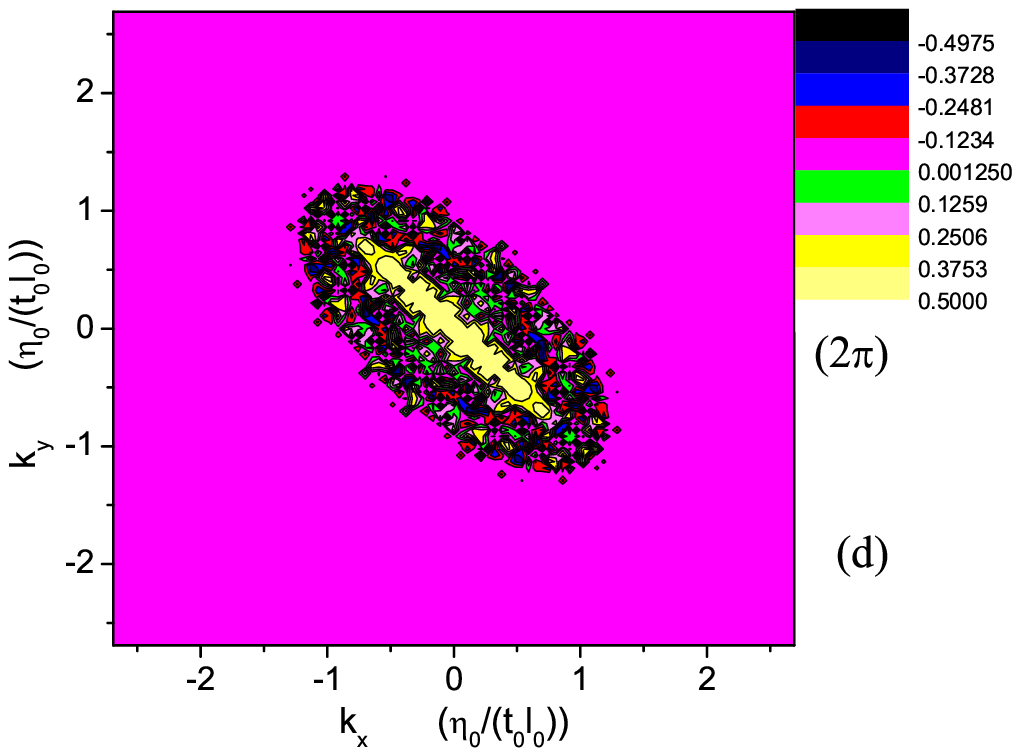} \caption{(Color online) Contour plot of
geometric phase acquired by fermions in valley ``$+$" and in the
upper band as a function of $k_x$ and $k_y$. The parameters for
vibrations are $\eta_x=\eta_y=\eta_0=0.078t_0$, $\hbar
\omega=0.0037t_0$, $\alpha_x=0$, and $\alpha_y = 0.6$.} \label{fig5}
\end{figure}

In summary, we investigate the geometric phase and the change of
dispersion relation of fermion states under relative vibrations of
two sublattices in graphene. For a circular vibration there are two
regions in the momentum space: in one region the geometric phase
acquired by the fermion states can be approximated by the adiabatic
Berry phase and the average energy spacing between two bands is
enlarged; in the other region the geometric phase exhibits random
oscillations in changing the momentum, much different from the
adiabatic value, and the average energy spacing shrinks. The energy
uncertainty of fermions due to vibrations shows spikes in the
chaotic region. For elliptic paths of vibrations the distribution of
the adiabatic and chaotic regions crucially depends on the path
shapes. The results suggest a possible dephasing mechanism which
causes classical-like transport properties in graphene.

 \acknowledgments{This work was supported by the State
Key Programs for Basic Research of China (2005CB623605 and
2006CB921803), and by National Foundation of Natural Science in
China Grant Nos. 10474033, \revision{10704040,} and 60676056.}


\end{document}